\documentclass[aps,prb,10pt, twocolumn,groupedaddress,showpacs,superscriptaddress,amssymb,amsmath]{revtex4-2}
\usepackage{physics}
\usepackage{slashed}
\usepackage{graphicx}
\newcommand{\be}{\begin{equation}}
	\newcommand{\ee}{\end{equation}}

\usepackage{mathrsfs}

\usepackage{ucs}
\usepackage{soul}
\usepackage[utf8x]{inputenc}
\usepackage{amsmath,bm}
\usepackage{amsfonts}
\usepackage{comment}
\usepackage{amssymb}
\usepackage{array}
\usepackage{slashed}
\usepackage{lipsum}
\usepackage{hyperref}
\hypersetup{
	colorlinks=true,
	linkcolor=blue,
	citecolor=cyan
}

\newcolumntype{C}[1]{>{\centering\let\newline\\\arraybackslash\hspace{0 pt}}m{#1}}

\begin{document}
	\title{Statistical Correlators and Tripartite Entanglement }
 \author{Sakil Khan}
\email{sakil.khan@students.iiserpune.ac.in}
\affiliation{Department of Physics,
		Indian Institute of Science Education and Research, Pune 411008, India}
\author{ Dipankar Home}
\email{quantumhome80@gmail.com}
\affiliation{Center for Astroparticle Physics and Space Science (CAPSS), Bose Institute, Kolkata-700 091, India}
\author{ Urbasi Sinha}
\email{
usinha@rri.res.in}
\affiliation{ Raman Research Institute, C. V. Raman Avenue, Sadashivanagar, Bengaluru, Karnataka 560080, India }
  \author{Sachin Jain}
\email{sachin.jain@iiserpune.ac.in}
\affiliation{Department of Physics,
		Indian Institute of Science Education and Research, Pune 411008, India}
	
	\date{\today}
	
	\begin{abstract}
		It has recently been argued that among the various suggested measures of tripartite entanglement, the two particular measures, viz. the Concurrence Fill and the Genuine Multipartite Concurrence are the only 'genuine' tripartite entanglement measures based on certain suitably specified criteria.  In this context,  we show that these two genuine tripartite entanglement measures can be empirically determined for the two important classes of tripartite entangled states, viz. the generalized GHZ and the generalized W states using the derived relationships of these two measures with the observable statistical correlators like the Pearson correlator and mutual information. Such a formulated scheme would therefore provide for the first time the means to exactly quantify tripartite entanglement, crucial for the proper assessment of its efficacy as resource. We also point out two specific applications of this scheme, viz. a) Enabling empirical demonstration of the potentially significant feature of inequivalence between Concurrence Fill and Genuine Multipartite Concurrence in quantitatively assessing which of the two given tripartite states is more entangled than the other one. b) Enabling experimental detection of the recently predicted phenomenon of entanglement sudden death for a tripartite system.
	\end{abstract}
	\maketitle  
	
\section{Introduction}
Entanglement is one of the key fundamental features of quantum mechanics. Apart from having deep-seated conceptual implications, it serves as a crucial resource for wide-ranging applications in the rapidly developing field of quantum information processing and quantum computation \cite{PhysRevLett.70.1895,ieee1984proceedings,nielsen_chuang_2010, PhysRevLett.70.1895, PhysRevA.58.4394}. In this context, it is important to quantitatively assess in an operationally useful way the efficacy of a given entangled state for any specified application. For this purpose, it is then necessary to suitably quantify how much entangled is any given state. Of course, using the tomographic technique, one can exactly determine any given state, but this procedure requires an increasing number of measurements as the state dimension increases. Thus, considering any arbitrary dimensional entangled state, for practical usefulness, it is desirable that the required quantification of entanglement is realized in terms of a minimal number of measurements. Now, we note that the usual applications of entanglement have essentially used bipartite entangled states. In this context, the quantification of entanglement has largely been analyzed in terms of the extensively studied entanglement measures like Negativity (N) and Entanglement of Formation (EOF)  \cite{nat1, PhysRevA.97.042338, Erker2017quantifyinghigh, PhysRevA.95.042340, PhysRevLett.118.110501}.  However, it is important to stress that all the earlier studies concerning the quantification of bipartite arbitrary dimensional entanglement, using a limited number of measurements, have essentially focused on providing tighter lower bounds on entanglement measures like N and EOF \cite{nat1, PhysRevA.97.042338, Erker2017quantifyinghigh, PhysRevA.95.042340, PhysRevLett.118.110501}.


It is only very recently a general scheme has been developed for actually determining the entanglement measures for arbitrary dimensional bipartite entangled states by directly relating them to observable quantities 
\cite{macc,sinha, Ghosh_2022,2022arXiv220106188S}.
In particular, this has enabled experimental determination of bipartite entanglement measures for a prepared two qutrits pure photonic state using the analytically derived relationships with statistical correlators like Pearson Correlation Coefficient (PCC), Mutual Predictability (MP), and Mutual Information (MI).
It is these works, formulating and applying such a general scheme for quantifying bipartite arbitrary dimensional entanglement through the use of statistical correlators for evaluating the entanglement measures that motivate our present paper. More precisely, here we probe the question  whether such a scheme can be extended for tripartite entangled states. This underpinning motivation is further strengthened by the following considerations.

Although the characterization of multipartite entanglement using suitably defined entanglement measures is not a straightforward extension of the scheme employed for the bipartite case, a number of studies have been made over the last two decades proposing a variety of multipartite entanglement measures. However, only recently, considering in particular the tripartite qubit case, the significance of defining what has been called a genuine multipartite entanglement (GME) measure has been clearly brought out through the work by Xie and Eberly \cite{triangle, Xie2}. They have argued that for faithfully quantifying the tripartite entanglement as a useful resource for potential applications, an appropriately defined GME needs to be used satisfying the conditions that (a) the measure has to be zero for all product and biseparable states, (b) the measure has to be positive for all nonbiseparable states, which turn out to be essentially GHZ and W classes of states in the tripartite qubit case since it was earlier pointed out \cite{PhysRevA.62.062314} that all three-qubit states can be separated into four distinct classes: product states, biseparable states, the GHZ class, and the W class.

One such tripartite qubit entanglement measure satisfying the above conditions for GME has been proposed by Xie and Eberly \cite{triangle} called Concurrence Fill (CF). Further, by analyzing all other tripartite qubit entanglement measures proposed to date, they have pointed out that the only other tripartite qubit entanglement measure that satisfies the conditions for GME is the one that was suggested by Ma et al. \cite{PhysRevA.83.062325} and has been called genuine multipartite concurrence (GMC) \footnote{For the three-qubit system, Generalized Geometric Measure (GGM) \cite{PhysRevA.95.022301, PhysRevA.81.012308} is equivalent to GMC.}. In our present paper, our treatment will be based on these two genuine tripartite entanglement measures.

Earlier works had shown the usefulness of tripartite entanglement for studying nonlocality, for example, experimental test for nonlocality has been performed using GHZ state, as well as the potentiality of tripartite qubit entanglement for applications such as in teleportation \cite{PhysRevLett.70.1895, PhysRevA.58.4394, ZOU201776, JOO2002324, SHI2002161, P1} and its less vulnerability to cheating than the bipartite entanglement based teleportation.
Other important applications of tripartite entanglement have also been studied such as quantum dense coding \cite{Nie, ROY20181709, D1, D2, YE2005330}, quantum cryptography \cite{PhysRevA.59.1829,2007PhLA..360..518B, JIN200667, Man2006, CaoHaiJing2006}.
Moreover, very recently, the possibility of tripartite entanglement being used even for demonstrating the phenomenon of entanglement sudden death (ESD) has been revealed, crucially by using CF and GMC, thereby yielding interesting insights into the type of initial condition that can give rise to ESD \cite{2022arXiv221001854X}.

Against the above background,
given the various potential applications of tripartite entangled states, the exact quantification of the entanglement of such states by determining the relevant entanglement measures in order to precisely assess the efficacy of their uses is becoming increasingly important. In this context,
 in the present paper, we will not only show the way the statistical correlators PCC and MI  (appropriately defined for the tripartite entangled qubit systems ) can be related to the genuine tripartite entanglement measures, viz. CF and GMC for the GHZ and W classes of states, thereby enabling their empirical determination, our treatment will have the following two significant implications, too.

a) Xie and Eberly had given an example to illustrate an important feature that the genuine tripartite qubit entanglement measures CF and GMC are inequivalent in the sense of not being monotonic with respect to each other. Here we may recall that while it is well known that for the bipartite qubit entangled states, the EMs are monotonically related, for the bipartite higher dimensional entangled states, it is only recently the inequivalence between the EMs for such states has been probed in detail, and the scheme by which such inequivalence can be experimentally demonstrated has been concretely outlined in terms of measurements of statistical correlators with respect to an experimental architecture that has been developed using bipartite qubit entangled system. Motivated by this line of study, we have discussed in our present paper how the scheme formulated here for the tripartite entangled GHZ and W states relating CF and GMC with the observable statistical correlators PCC and MI can be used for empirical demonstration of the inequivalence between CF and GMC, provided one can develop an appropriate experimental architecture. Such inequivalence can have interesting fundamental as well as practical implications, and hence its empirical demonstration would be significant.

b) Further, we have related PCC, MI, and Mutual Predictability (MP) with the GMC of an initial GHZ state which is time evolving through an amplitude damping Markovian Channel towards showing sudden death of tripartite entanglement. This result can be used for any future effort to study empirically ESD for the tripartite case.
Finally, we would like to mention that, although our treatment is based on using essentially pure tripartite entangled states, we have also indicated briefly in the end, the way this scheme can be extended for mixed states by obtaining an illustrative relationship between CF and MI for a mixture of GHZ and W states. 

The rest of the paper is organized as follows.  In section \ref{sec2} we provide the relevant background which will help us to define PCC and MI for the tripartite system. In section \ref{sec3} we review the genuine measures of tripartite entanglement and in section \ref{rsc} we review the statistical correlators in the bipartite system. In section \ref{sec4} we generalize the definition of statistical correlators in the tripartite system. In section \ref{sec5} we derive  monotonic relation between statistical correlators and genuine measure of tripartite entanglement for a few classes of tripartite state. Using these monotonic relations we can determine the genuine tripartite entanglement measure. In section \ref{sec6} we provide a scheme to demonstrate the inequivalence between different measures of tripartite entanglement empirically. In section \ref{sec7} we show how one can experimentally detect the sudden death of tripartite entanglement using statistical correlators. In the discussion section \ref{sec9}, we give a summary of our work as well as some future directions. 
For completeness, we have also included three Appendices. In Appendix \ref{ngm}, we have discussed two widely used measures of tripartite entanglement which are not genuine measures, viz. Global Measure of Entanglement ($ G_{123}$) and  Tangle Measure of Entanglement ($ \tau_{123}$). In the Appendix \ref{tmc},
we extend the proof of the Maccone conjecture \cite{macc} for the tripartite system and verify it with two examples.
 In Appendix \ref{sec8} we extend our study for the mixed tripartite state.

\section{Background} \label{sec2}

\subsection{Review  of genuine measures of tripartite entanglement} \label{sec3}
According to the criteria justified in \cite{triangle}, if a multipartite entanglement (ME) measure satisfies the following two  conditions
(a) the measure
is zero for all product and biseparable states
(b) the measure is positive for all nonbiseparable
states
 then the measure is called a genuine
multipartite entanglement (GME) measure. In this subsection, we are going to review two genuine measures of tripartite entanglement which are  Concurrence Fill (CF) and Genuine Multipartite Concurrence (GMC) following \cite{PhysRevA.83.062325, triangle}.
 In \cite{triangle}, the authors have shown that
these two measures satisfy all the properties to become a genuine measure of entanglement. We will explicitly calculate  CF and GMC for the generalized $\ket{GHZ} $
state and generalized $\ket{W}  $ state. 
These states are defined as 
 \begin{equation} \label{eqghz}
	\begin{split}
		\ket{GHZ} = a\ket{000}+b\ket{111}\;,
	\end{split}
\end{equation}
\begin{equation} \label{eqw}
	\begin{split}
		\ket{W} = \cos{\theta}\ket{100}+\frac{\sin{\theta}}{\sqrt{2}}\ket{010}+\frac{\sin{\theta}}{\sqrt{2}}\ket{001}\;,
	\end{split}
\end{equation}
where $a, b \in [0,1]$ and $\theta \in [0,\pi/2]$.
In appendix \ref{ngm}, we discussed two other well known measures of tripartite entanglement, the Global Measure of Entanglement ($ G_{123}$) and  Tangle Measure of Entanglement ($ \tau_{123}$). However, as argued in \cite{triangle}, these are not
   genuine measures of tripartite entanglement. 
\subsubsection{Concurrence Fill (CF)}
\begin{figure}
	\includegraphics[scale=.7]{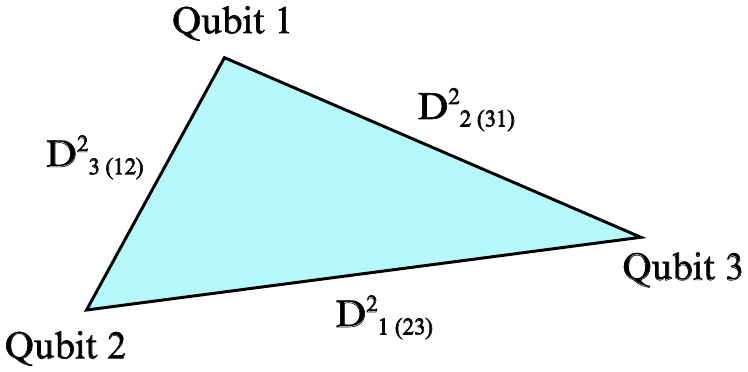}
	\caption{This figure illustrates the concurrence triangle for a three-qubit system. The edges of the triangle, $D^2_{i(jk)}$, 
 are equal to the squares of the bipartite concurrences
 defined as $D^2_{i(jk)}=2\Big(1-Tr_{i}[Tr_{jk}\rho]^2\Big)$ . }
	\label{fig:0}
\end{figure}

CF for a tripartite state
 $\rho$ is given by the square root of the area of the concurrence triangle \cite{triangle} shown in Fig.\ref{fig:0} and is given by
\begin{equation}\label{eqf}
	F_{123}(\rho)= \Big[\frac{16}{3}Q(Q-D^2_{1(23)})(Q-D^2_{2(13)})(Q-D^2_{3(12)})\Big]^{\frac{1}{4}}\;,
\end{equation}
where $Q=\frac{1}{2}(D^2_{1(23)}+D^2_{2(13)}+D^2_{3(12)})$ and
\begin{equation}\label{eqdijk}
D_{i(jk)}=\sqrt{2\Big(1-Tr_{i}[Tr_{jk}\rho]^2\Big)}
\end{equation}
 with $i = 1, 2, 3$ and $jk =
23, 13, 12$ respectively. Let us illustrate this using the generalized $\ket{GHZ} $ and generalized $ \ket{W}$ state.

For the generalized $\ket{GHZ}$ state defined in Eq.\eqref{eqghz}, 
the Concurrence Fill ($F_{123}$) 
is \cite{triangle}
\begin{equation}\label{eqghzf}
	F_{123}=4a^2b^2.
\end{equation}
Here we have used the fact that 
all three edges of the triangle, as shown in Fig.\ref{fig:0}, are given by
\begin{equation}\label{eqghze}
	D^2_{1(23)}=D^2_{2(31)}=D^2_{3(12)}=4a^2b^2.
\end{equation}
Similarly, we calculate all three edges of the triangle shown in Fig.\ref{fig:0} for the generalized $\ket{W}$ state, defined in Eq.\eqref{eqw}, 
 and they are \cite{triangle} 
\begin{align}\label{eqwe}
	&D^2_{2(31)}=D^2_{3(12)}=\sin^2{\theta}\big(1+\cos^2{\theta}\big),
	\nonumber\\
	&D^2_{1(23)}=\sin^2{2\theta}.
\end{align}
The  CF ($F_{123}$) for the generalized $\ket{W}$ state  
is 
\begin{equation}\label{eqwf}
	F_{123}=\sin{2\theta}\big[\frac{4}{3}\big(1-\cos^2{\theta}(1+\sin^2{\theta})\big)\big(1+\cos^2{\theta}(3\sin^2{\theta}-1)\big)]^{\frac{1}{4}}\;.    
\end{equation}
\subsubsection{Genuine Multipartite Concurrence (GMC)}
Let us now define the other genuine measure of tripartite entanglement i.e. Genuine Multipartite Concurrence  (GMC). 
For a 3-qubit system, the GMC ($C_{GMC}$) is given by the least edge of the triangle shown in Fig.\ref{fig:0} \footnote{In Ref. \cite{triangle}, GMC is denoted by $C_{GME}$ but here to avoid confusion we denoted it by $C_{GMC}$.} i.e. 
\begin{equation}\label{eqgmcdef}
  C_{GMC}= \text{least of } \{ D^2_{1(23)}, D^2_{2(13)}, D^2_{3(12)}\}\;,
\end{equation}
where $D^2_{i(jk)}$ is defined in Eq. \eqref{eqdijk}. Let us calculate GMC for the same two states.

For the generalized $\ket{GHZ}$ state defined in Eq.\eqref{eqghz}, we obtain the GMC using Eq.\eqref{eqghze} which is given by \cite{triangle}
\begin{equation}\label{eqghzgmc}
	C_{GMC}=F_{123}=4a^2b^2.
\end{equation}
Similarly, GMC for the generalized $\ket{W}$ state  is \cite{triangle}
\begin{align}\label{eqwgmc}
	C_{GMC}&=\sin^2{\theta}\big(1+\cos^2{\theta}\big);\;\;\;\;\;\;0 \leq \sin{\theta} \leq \sqrt{\frac{2}{3}} \nonumber\\
	&=\sin^2{2\theta};\;\;\;\;\;\;\;\;\;\; \;\;\;\;\;\;\;\;\;\;\;\sqrt{\frac{2}{3}}\leq \sin{\theta} \leq 1
\end{align}
where we have used Eq.\eqref{eqwe}.
\subsection{Review  of statistical correlators in the bipartite system}\label{rsc}
In this subsection, following \cite{macc,sinha} we define three different statistical correlators which are  Pearson Correlation Coefficient (PCC), Mutual Information (MI), and Mutual Predictability (MP). The PCC for the bipartite system for any two  observables $A$ and $B$ is defined as
\begin{equation} \label{eqpcc}
	\begin{split}
		C_{A B} = \frac{\langle AB \rangle-\langle A \rangle\langle B \rangle}{\sigma_{A}\sigma_{B}}\;,
	\end{split}
\end{equation}
where $\langle A \rangle=Tr[A\rho]$ is the expectation value of the operator $A$ on the state $\rho$, and $\sigma^2_{A}=\langle A^2 \rangle-\langle A \rangle^2$ is the variance. 

The MI of a $d\times d$ bipartite state $\rho$ with respect to the two
chosen bases $\{\ket{a_{i}
	, b_{j}} \}$ is defined as
\begin{equation} \label{eqmi}
	\begin{split}
		I_{A B} = \sum^{d,d}_{i,j=0}p(i,j)\log_{2}\bigg(\frac{p(i,j)}{p_{a}(i)p_{b}(j)}\bigg)\;,
	\end{split}
\end{equation}
where
\begin{equation}
	p(i,j)=\bra{a_{i}
		, b_{j}}\rho\ket{a_{i}
		, b_{j}}  \;,
\end{equation}
\begin{equation}
	p_{a}(i)=\bra{a_{i}}
	Tr_{b}\rho\ket{a_{i}}\;.
\end{equation}
One can easily generalise the definition of MI for $d_{1}\times d_{2}$ dimensional bipartite system in the following way
\begin{equation} \label{eq4}
	\begin{split}
		I_{A B} = \sum^{d_{1},d_{2}}_{i,j=0}p(i,j)\log_{2}\bigg(\frac{p(i,j)}{p_{a}(i)p_{b}(j)}\bigg)\;.
	\end{split}
\end{equation}
On the other hand, MP of a $d\times d$ bipartite state $\rho$ with respect to the two
chosen bases $\{\ket{a_{i}
	, b_{j}} \}$ is defined as
\begin{equation} \label{eqmpd}
	\begin{split}
		P_{A B} = \sum^{d}_{i=0}\bra{a_{i}
			, b_{i}}\rho\ket{a_{i}
			, b_{i}} \;.
	\end{split}
\end{equation}


\section{Results}
In this section, we provide the key results of our work. We begin by formulating the definition of statistical correlators in the tripartite system. We calculate these measures for specific examples and show that these are monotonic with respect to the genuine tripartite entanglement measures, which were discussed in the previous section.
We next demonstrate two applications of these monotonic relations namely showing inequivalence between different genuine measures of tripartite entanglement and the sudden death of tripartite entanglement.
\subsection{Extension of Statistical correlators in the tripartite system}\label{sec4}
 In this section,  we are going to generalize the definition of PCC, MI and MP for the tripartite system. 
 We know that every tripartite system can be thought of as a bipartite system by choosing any subsystem of the tripartite system as subsystem-1 and the rest of the tripartite system as subsystem-2. For the tripartite system, there can exist three such different bi-partitions and they are-
1-(23), 2-(13) and 3-(12). Let us start our discussion by defining PCC for the tripartite system.

\subsubsection{PCC for tripartite system}
 Let us consider the  bi-partition  1-(23),
and assume $A^1$ is an observable of subsystem-1 and similarly $B^{23}$ is an observable of subsystem-(23). Now let us define the following quantity
\begin{align}
&	C^{1-23}_{AB}= |C_{A^1B^{23}}| \;,
		\end{align}
 where $C_{A^1B^{23}}$ is the bipartite PCC defined in equation (\ref{eqpcc}) . Similarly, for the other two combinations, we can define the following quantities- $C^{2-13}_{AB}$ and $C^{3-12}_{AB}$.
We define the PCC in the tripartite system which we denoted by $	C^{123}_{AB} $ as the geometric mean of $C^{i-jk}_{AB}$ i.e.
\begin{equation}\label{eqspcc}
	C^{123}_{AB}=(C^{1-23}_{AB}C^{2-13}_{AB}C^{3-12}_{AB})^{\frac{1}{3}}\;.
\end{equation}
One of the motivations to define PCC in this way is that if any of the $C^{i-jk}_{AB}$ is zero then the PCC becomes zero. Note that, this way of generalizing PCC is not unique. Let us illustrate the use of definition in \eqref{eqspcc} with some examples.

\subsection*{Calculation of PCC for specific examples}
Generalized $\ket{GHZ}$ state and Generalized $\ket{W}$ state are two well known tripartite states. Below we calculate PCC for these states. 
\subsubsection*{a. Generalized $\ket{GHZ}$ state:}
The generalized  $\ket{GHZ}$ state is defined in \eqref{eqghz}.
Let's say we want to first calculate $C^{1-23}_{AB} $. To do that, we have to first fix the observable with respect to which we want to calculate this quantity.
Assume $A^1=\sigma_{X}$ is the observable acting on the subsystem-1  and 
 $B^{23}=\sigma_{X}\otimes\sigma_{X}$ is the observable acting on the subsystem-(23).
Similarly, to calculate $C^{2-13}_{AB} $ and $C^{3-12}_{AB} $, we choose the same observable i.e.   
\begin{align}
	&A^1=A^2=A^3=\sigma_{X}, B^{23}=B^{13}=B^{12}=\sigma_{X} \otimes \sigma_{X}\;.
\end{align}
Let's denote PCC ($ 	C^{123}_{AB}$) with respect to these observables by $	C^{123}_{X}$.
Using equation \eqref{eqspcc}, we can show that $	C^{123}_{X}$ is given by
\begin{equation}\label{eqcghz}
	C^{123}_{X}=2ab\;.
\end{equation}
Note that, in the limit $a=0$ or $b=0$ which corresponds to the separable state, $	C^{123}_{X}$ is zero. For $a=b=1/\sqrt{2}$ which corresponds to the $\ket{GHZ}$ state,  $	C^{123}_{X}$ is maximum and is given by 1.
\subsubsection*{b. Generalized $\ket{W}$ state:}
The generalized $\ket{W}$ state is defined in \eqref{eqw}.
For this state, let's assume $A^1=\ket{+}\bra{+}$ is the observable  acting on the subsystem-1 and 
$B^{23}=\ket{+}\bra{+}\otimes\ket{+}\bra{+}$ is the observable  acting on the subsystem-(23), where $\ket{+}=\frac{\ket{0}+\ket{1}}{\sqrt{2}} $.
Similarly, to calculate $C^{2-13}_{AB} $ and $C^{3-12}_{AB} $, we choose the same observable i.e.   
\begin{align}\label{eqwo2}
A^1=A^2=A^3=\ket{+}\bra{+}, B^{23}=B^{13}=B^{12}=\ket{+}\bra{+} \otimes \ket{+}\bra{+}\;.
\end{align}
Let's denote PCC with respect to these observables by $C^{123}_{+}$.
Using Eq.\eqref{eqspcc}, we can show that $	C^{123}_{+}$ is given by
\begin{widetext}
\begin{equation}\label{eqwc1}
	C^{123}_{+}=\bigg[\bigg(\frac{\sqrt{2}\sin{2\theta}}{\sqrt{(1+\sin^2{\theta})(2+\cos^2{\theta})}}\bigg).\bigg(\frac{\sin{2\theta}+\sqrt{2}\sin^2{\theta}}{\sqrt{6+2\sqrt{2}\sin{2\theta}-\sin^2{2\theta}}}\bigg)^2\bigg]^{\frac{1}{3}}   
\end{equation}
\end{widetext}
Note that, $C^{123}_{+}$ is zero in the limit $\theta=0,\pi/2$ which corresponds to the separable state and bi-separable state respectively. On the other hand for $\theta=\arctan{\sqrt{2}}$ which corresponds to the $\ket{W}$ state,  $C^{123}_{+}$ is maximum and is given by 0.676 .

\subsubsection{MI for tripartite system}
We now turn our attention to defining MI for the tripartite system.
As in the case of PCC, let us denote the MI for the bi-partition $1-(23)$ by $I^{1-23}_{A B}$ and using the Eq.\eqref{eq4} we can calculate $I^{1-23}_{A B}$.
Similarly, for the other two bi-partitions, we define  $I^{2-13}_{A B}$ and $I^{3-12}_{A B}$.
As in the case of PCC, we define  MI for the tripartite system as the geometric mean of $I^{i-jk}_{A B}$,
\begin{equation}\label{eqspcca}
	I^{123}_{A B}=\big(I^{1-23}_{A B}I^{2-13}_{A B}I^{3-12}_{A B}\big)^{\frac{1}{3}}
\end{equation}

\subsection*{Calculation of MI for specific examples}
 Below we calculate MI for the generalized $\ket{GHZ}$ state and generalized $\ket{W}$ state.
 \subsubsection*{a. Generalized $\ket{GHZ}$ state:}
We denoted $I^{1-23}_{A B}$ with respect to the eigenbasis of $\sigma_{X}\otimes(\sigma_{X}\otimes\sigma_{X})$ by $I^{1-23}_{X}$ and for generalized $\ket{GHZ}$ state defined in \eqref{eqghz} it is given by
\begin{equation}
	I^{1-23}_{X}= \bigg(\frac{1+2ab}{2}\bigg)\log_{2}(1+2ab)+\bigg(\frac{1-2ab}{2}\bigg)\log_{2}(1-2ab)\;.
\end{equation}
Similarly, we have calculated $I^{2-13}_{X}$ and $I^{3-12}_{X}$ and they are equal to $I^{1-23}_{X}$ i.e. $I^{2-13}_{X}=I^{3-12}_{X}=I^{1-23}_{X}$.
Using equation \eqref{eqspcca}, we get 
\begin{equation}\label{eqighza}
	I^{123}_{X}=\bigg(\frac{1+2ab}{2}\bigg)\log_{2}(1+2ab)+\bigg(\frac{1-2ab}{2}\bigg)\log_{2}(1-2ab).
\end{equation}
Note that, in the limit $a=0$ or $b=0$ which corresponds to the separable state, $	I^{123}_{X}$ is zero. Whereas for $a=b=1/\sqrt{2}$ which corresponds to the $\ket{GHZ}$ state,  $	I^{123}_{X}$ is maximum and is given by 1. \\

\subsubsection*{b. Generalized $\ket{W}$ state:}
For the generalized $\ket{W}$ state defined in  Eq.\eqref{eqw}, we obtain the following results
\begin{equation}
	I^{1-23}_{X}= \frac{m^2}{4}\log_{2}\bigg(\frac{m^2}{1+\sin^2{\theta}}\bigg)+\frac{n^2}{4}\log_{2}\bigg(\frac{n^2}{1+\sin^2{\theta}}\bigg)\;,
\end{equation}
\begin{align}
	I^{2-13}_{X}=&I^{3-12}_{X} =\frac{m^2}{4}\log_{2}\bigg(\frac{2m^2}{2+\sqrt{2}\sin{2\theta}}\bigg)\nonumber\\
	&+\frac{n^2}{4}\log_{2}\bigg(\frac{2n^2}{2-\sqrt{2}\sin{2\theta}}\bigg)\nonumber \\
	&+\frac{\cos^4{\theta}}{4}\log_{2}\bigg(\frac{4\cos^4{\theta}}{(2+\sqrt{2}\sin{2\theta})(2-\sqrt{2}\sin{2\theta})}\bigg)\;,
\end{align}
where
\begin{align}
	m=(\cos{\theta}+\sqrt{2}\sin{\theta}) \;\;\;\text{and }\;\;\; n=(\cos{\theta}-\sqrt{2}\sin{\theta})\;.
\end{align}
Using equation \eqref{eqspcca}, we get
\begin{equation}\label{eqiw}
	I^{123}_{X}=\bigg(I^{1-23}_{X}\;(I^{2-13}_{X})^2\bigg)^{\frac{1}{3}}\;.
\end{equation}
Similarly, we  calculate   $I^{123}_{Z}$ and $I^{123}_{Y}$ for this state and  they are given by
\begin{align}\label{eqizw}
	&I^{123}_{Z}=\bigg(I^{1-23}_{Z}\;(I^{2-31}_{Z})^2\bigg)^{\frac{1}{3}} \;\;\;\;\;\text{and}\nonumber\\
	& I^{123}_{Y} =I^{123}_{X}\;,
\end{align}
where
\begin{align}
	I^{1-23}_{Z} =&-\cos^2{\theta}\log_{2}(cos^2{\theta})-\sin^2{\theta}\log_{2}(sin^2{\theta})\;,
\end{align}
\begin{align}
	I^{2-13}_{Z} =&\cos^2{\theta}\log_{2}(\frac{4}{3+\cos{2\theta}})\nonumber\\
	&+\frac{sin^2{\theta}}{2}\bigg(\log_{2}(\frac{4}{3+\cos{2\theta}})-\log_{2}(\frac{sin^2{\theta}}{2})\bigg)\;.
\end{align}
Note that, the limit $\theta=0$ and $\theta=\pi/2$ which corresponds to the separable state and the bi-separable state respectively, $	I^{123}_{Z}$ is zero. For $\theta=\arctan{\sqrt{2}}$ which corresponds to the $\ket{W}$ state,  $	I^{123}_{Z}$ is maximum and is given by 0.840.
\subsubsection{MP for tripartite system}
Mutual Predictability (MP) of a $d\times d$ dimensional bipartite system is defined in Eq.\eqref{eqmpd}. We can generalize the definition of MP in the $d\times d\times d$ dimensional tripartite system with respect to the three
chosen bases $\{\ket{a_{i}
	, b_{j},c_{k}} \}$ as  
\begin{equation} \label{eqmptd}
	\begin{split}
		P_{A B C} = \sum^{d}_{i=0}\bra{a_{i}
			, b_{i}, c_{i}}\rho\ket{a_{i}
			, b_{i},c_{i}} \;.
	\end{split}
\end{equation}

\subsection*{Calculation of MP for specific examples}
We are going to calculate MP for the generalized $\ket{GHZ}$ state and generalized $\ket{W}$ state. 
To compute MP for the generalized $\ket{GHZ}$ state defined in \eqref{eqghz} , we choose  eigenbasis of  $\sigma_{Z}\otimes(\sigma_{Z}\otimes\sigma_{Z}).$ Denoting MP by $P_{Z},$  we obtain
\begin{equation}\label{eqpzghz}
	P_{Z}= 1\;.
\end{equation}
Note that, for the generalized GHZ state  $	P_{Z}$ is constant. 
For the generalized $\ket{W}$ state defined in  Eq.\eqref{eqw}, we obtained the following result
\begin{equation}\label{eqpxw}
	P_{X}= \frac{1}{8}\Big( 3-\cos{2 \theta }+2\sqrt{2}\sin{2\theta}\Big)\;.
\end{equation}
Note that, to compute MP for the generalized $\ket{W}$ state, we choose  eigenbasis of  $\sigma_{X}\otimes(\sigma_{X}\otimes\sigma_{X})$. Here, we denote MP by $P_{X}$.

\subsection{Relation between statistical correlators and genuine measures of tripartite entanglement } \label{sec5}
In this section, we demonstrate an explicit relation between statistical correlators defined in the previous section and genuine measures of tripartite entanglement. Using these relations in principle one can experimentally measure tripartite entanglement measures like concurrence fill, GMC, etc. 

\subsubsection{Relation between  CF, GMC, and  PCC}
In this subsection, we relate  CF and GMC with PCC for the generalized GHZ state and generalized W state.
We got a monotonic relation between them which we can use to experimentally detect CF and GMC. Below we give the explicit calculations.
\subsubsection*{a. Generalized $\ket{GHZ}$ state:}
We have the explicit form of $  C^{123}_{X}$, $ C_{GMC}$ and $F_{123}$ for this state, so by comparing them, we can get the relation between them. By comparing  Eq.\eqref{eqcghz}, Eq.\eqref{eqghzf} and Eq.\eqref{eqghzgmc} we get the following analytical relation between the concurrence fill ($F_{123}$), GMC and PCC ($C^{123}_{X}$),
\begin{align}\label{eqcf}
     &C^{123}_{X}=\sqrt{F_{123}},\nonumber\\
     &C^{123}_{X}=\sqrt{C_{GMC}}.
\end{align}
Encouraged by the above results, we next study for the generalized W states, the relationship of $C^{123}_{AB}$ with $F_{123}$ and $C_{GMC}$ respectively.
\subsubsection*{b. Generalized $\ket{W}$ state:}
\begin{figure}
    \includegraphics[scale=.7]{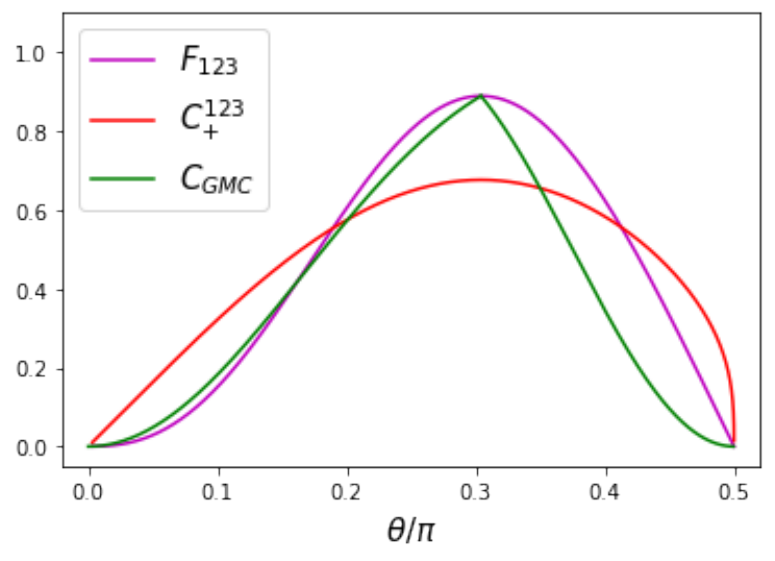}
    \caption{ We have plotted PCC ($C^{123}_{AB}$),  GMC ($C_{GMC}$) and CF ($F_{123}$) with the free parameter $\theta$ for the generalized W state.
    	$	C^{123}_{AB}$ with respect to the observable defined in Eq.\eqref{eqwo2} is denoted by $C^{123}_{+}$. All the functions have the same functional behavior i.e. they all first increase with $\theta$ and reach their maximum value and then decay. The maxima of $F_{123}$, $C_{GMC}$ and $C^{123}_{+}$ occurs at the same value of $\theta$ which is, $\theta/\pi=0.304$.}
    \label{fig:20}
\end{figure}

We have explicitly calculated $C^{123}_{+}$, $F_{123}$ and $C_{GMC}$ for the generalized $\ket{W}$ state in Eq.\eqref{eqwc1}, Eq.\eqref{eqwf} and Eq.\eqref{eqwgmc} respectively. In this case, we find that while $C^{123}_{+}$ cannot be analytically linked with either $F_{123}$ or $C_{GMC}$, we can study their relationship numerically.
 We have plotted $C^{123}_{+}$,  $C_{GMC}$ and $F_{123}$ with the free parameter $\theta$ in Fig.\ref{fig:20}.
From Fig.\ref{fig:20}, we can see all the functions have the same functional behavior i.e. they all first increase with $\theta$ and reach their maximum value and then decay. The maxima of $F_{123}$, $C_{GMC}$ and $C^{123}_{+}$ occurs at the same value of $\theta$ which is, $\theta/\pi=0.304$.
So it is evident from the plot that, we can use PCC to find  $F_{123}$ and $C_{GMC}$.


\subsubsection{Relation between  CF, GMC, and  MI}
In this subsection, we show relations between CF, GMC, and MI for the generalized GHZ state and generalized W state.
We obtain monotonic relation between them which we can use to experimentally detect CF and GMC. Below we give the explicit calculations.
\subsubsection*{a. Generalized $\ket{GHZ}$ state:}
\begin{figure}[h!]
	\includegraphics[scale=.63]{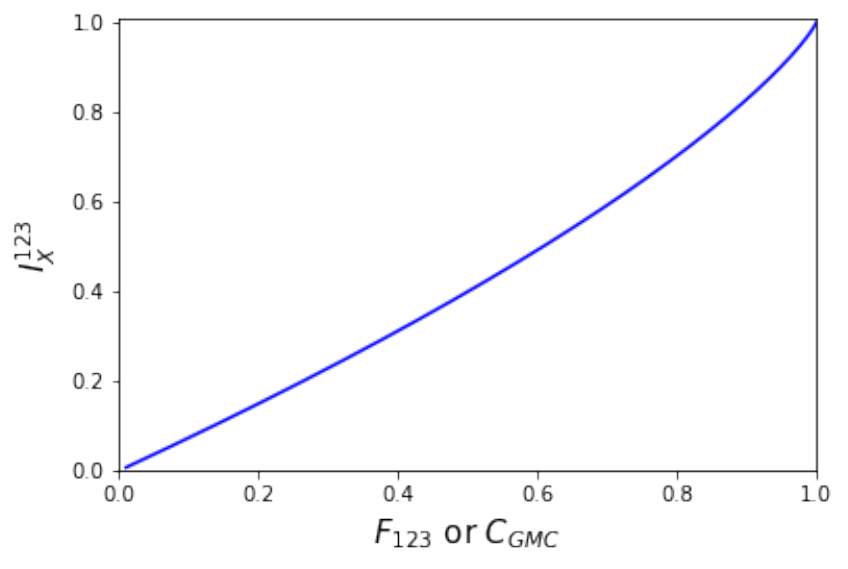}
	\caption{We have plotted MI ($I^{123}_{AB}$) vs CF ($F_{123}$) or GMC ($C_{GMC}$) for the generalized GHZ state. $	I^{123}_{AB}$ with respect to the $\sigma_{X}\otimes\sigma_{X}\otimes\sigma_{X}$ operator is denoted by $I^{123}_{X}$.
  From the plot, it is clear that we obtained a monotonic relation between $I^{123}_{X}$ and  $F_{123}$ or $C_{GMC}$. So by measuring $I^{123}_{X}$ we can find  $F_{123}$ or $C_{GMC}$.}
	\label{fig:003}
\end{figure}

For this state, we have the explicit form of  $F_{123}$,  $  C_{GMC}$ and $  I^{123}_{X}$ in Eq.\eqref{eqghzf}, Eq.\eqref{eqghzgmc} and  Eq.\eqref{eqighza} respectively. 
Using these results we get the following relations 
\begin{align}\label{eq12}
	I^{123}_{X}=&\bigg(\frac{1+\sqrt{F_{123}}}{2}\bigg)\log_{2}(1+\sqrt{F_{123}})\nonumber\\
	&+\bigg(\frac{1-\sqrt{F_{123}}}{2}\bigg)\log_{2}(1-\sqrt{F_{123}})
\end{align}
\begin{align}\label{eq12a}
	I^{123}_{X}=&\bigg(\frac{1+\sqrt{C_{GMC}}}{2}\bigg)\log_{2}(1+\sqrt{C_{GMC}})\nonumber\\
	&+\bigg(\frac{1-\sqrt{C_{GMC}}}{2}\bigg)\log_{2}(1-\sqrt{C_{GMC}})
\end{align}
Notice that we get the same analytical relation between MI and $F_{123}$ and MI and $C_{GMC}$ since for this state,  $F_{123}= C_{GMC}$, see \eqref{eqghzf} and \eqref{eqwgmc}. We have plotted Eq.\eqref{eq12} or Eq.\eqref{eq12a} in Fig.\ref{fig:003}. From the plot, it is clear that we got a monotonic relation between $I^{123}_{X}$ and  $F_{123}$ or $C_{GMC}$. So by measuring $I^{123}_{X}$ we can find $F_{123}$ or $C_{GMC}$.
\subsubsection*{b. Generalized $\ket{W}$ state:}
\begin{figure}
	\includegraphics[scale=.65]{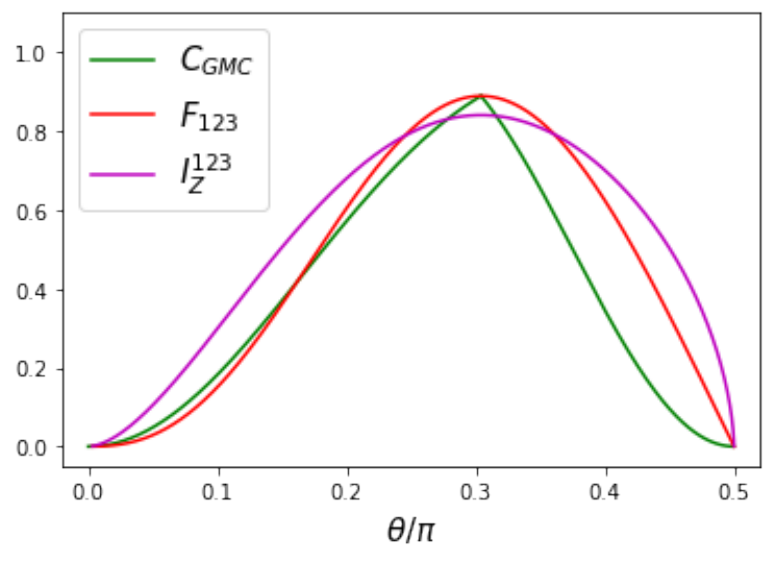}
	\caption{We have plotted $I^{123}_{AB}$,  $C_{GMC}$ and $F_{123}$ with the free parameter $\theta$ for the generalized W state.
		$	I^{123}_{AB}$ with respect to the $\sigma_{Z}\otimes\sigma_{Z}\otimes\sigma_{Z}$ operator is denoted by $I^{123}_{Z}$.
		From the plot, we can see that the mutual information $I^{123}_{Z}$ is monotonically related to GMC and triangle measure.  $I^{123}_{Z}$,  $F_{123}$ and $C_{GMC}$ have the same functional behavior with $\theta$. The maxima of $F_{123}$, $C_{GMC}$ and $I^{123}_{Z}$ occur at the same value of $\theta$ which is, $\theta/\pi=0.304$.}
	\label{fig:6}
\end{figure}

We have explicitly calculated the  CF ($F_{123}$), GMC and MI ($I^{123}_{AB}$) for the generalized $\ket{W}$ state in Eq.\eqref{eqwf}, Eq.\eqref{eqwgmc}  and Eq.\eqref{eqiw}-Eq.\eqref{eqizw} respectively. We have plotted $I^{123}_{Z}$,  GMC and $F_{123}$ with the free parameter $\theta$ in Fig.\ref{fig:6}. From the plot, we can see that the mutual information $I^{123}_{Z}$ is monotonically related to GMC and triangle measure.  $I^{123}_{Z}$,  $F_{123}$ and $C_{GMC}$ has the same functional behavior with $\theta$. So using $I^{123}_{Z}$, we can experimentally detect  $F_{123}$ and $C_{GMC}$.
\subsubsection{Relation between  CF, GMC, and  MP
}
We have explicitly calculated MP for the generalized GHZ state and generalized W state 
in Eq.\eqref{eqpzghz} and Eq.\eqref{eqpxw}. But they are not monotonic with genuine measures of tripartite entanglement calculated in \ref{sec3}.


\subsection{Two practical applications of monotonic relation between statistical correlators and genuine measures of tripartite entanglement}
In this section, we investigate the implications of monotonic relation between statistical correlators and genuine measures of tripartite entanglement. More precisely, we demonstrate two properties of tripartite entanglement namely inequivalence between different genuine measures of tripartite entanglement and sudden death of tripartite entanglement.

\subsubsection{Inequivalence between different measure of tripartite entanglement } \label{sec6}
\begin{figure}
	\includegraphics[scale=.6]{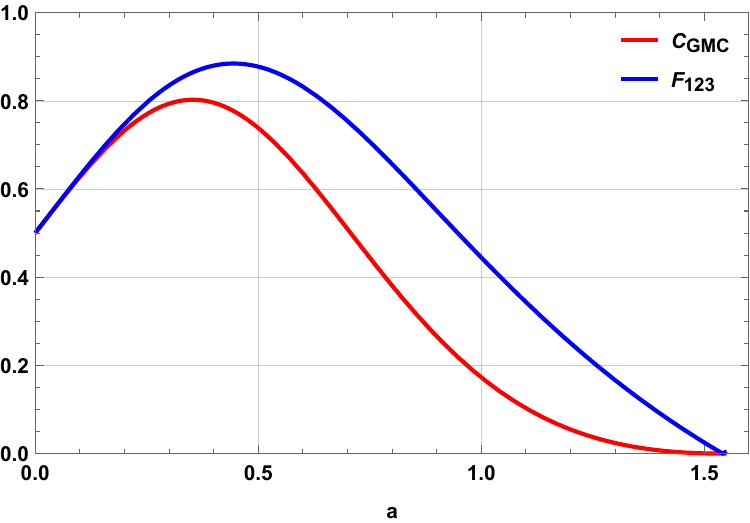}
	\caption{From the plot it is clear that $C_{GMC}$ and $F_{123}$ is not monotonically related to each other.The maxima of $C_{GMC}$ and
		the maxima of
		$F_{123}$ occurs at different $a$.}
	\label{fig:8}
\end{figure}

Inequivalence is an important aspect of entanglement measures. 
In this section, we discuss inequivalence between genuine measures of tripartite entanglement.
To understand the meaning of inequivalence between different measures, let's consider two entangled states, $\ket{\psi_{1}}$ and $\ket{\psi_{2}}$.  
Assume X and Y are two genuine measures of entanglement. If according to measure X  $\ket{\psi_{1}}$ is more entangled than $\ket{\psi_{2}}$ i.e. $X(\ket{\psi_{1}}) >X(\ket{\psi_{2}})$ but 
measure Y indicates $\ket{\psi_{2}}$ is more entangled than $\ket{\psi_{1}}$ i.e. $Y(\ket{\psi_{2}}) >Y(\ket{\psi_{1}})$ then 
we say X and Y are two inequivalent measures of entanglement.
 For the bipartite higher dimensional entangled states, only recently the inequivalence between the EMs has been probed in detail \cite{Ghosh_2022}.
 For the tripartite system, in Ref.\cite{triangle}, Xie and Eberly have given only one particular example to illustrate this inequivalence. They have shown that the entanglement measures CF and GMC are inequivalent in the sense of not being monotonic with respect to each other.
 More explicitly, they considered these two following states 
 \begin{align}
	&\ket{\psi_{1}} = \frac{1}{\sqrt{2}} \sin{(\pi/5)}\ket{000}+\frac{1}{\sqrt{2}} \cos{(\pi/5)}\ket{100}+\frac{1}{\sqrt{2}} \ket{111}\;,\nonumber\\
		&\ket{\psi_{2}} = \sin{(\pi/8)}\ket{000}+\sin{(\pi/8)}\ket{111}\;.
\end{align}
Now, in terms of GMC,
 $\ket{\psi_{2}}$ is more entangled than $\ket{\psi_{1}}$, since
$C_{GMC}(\ket{\psi_{2}})=0.5 >C_{GMC}(\ket{\psi_{1}}= 0.345$. However, in terms of CF,
$F_{123}(\ket{\psi_{2}})=0.5 <F_{123}(\ket{\psi_{1}}= 0.626$. But this is not a special example for which such inequivalence can be noticed. In fact, it is possible to construct quite  a number of  pairs of states 
 for which we can get such 
 inequivalence between GMC and CF. To show this emphatically, we proceed in the following way. We consider the following form of a general pure tripartite state involving an arbitrary parameter ‘a’ such that by choosing appropriate values of the parameter ‘a’ one can find a number of examples of pairs of states for which the entanglement measures GMC and CF are inequivalent in the sense explained earlier:
\begin{align} \label{eqx0}
	\ket{X} =&\frac{\bigg((\frac{a}{\sqrt{2}} + \cos( \pi/8) (1 - \sqrt{2} a)\bigg)\ket{000}}{\sqrt{N}}\nonumber\\
	&+\frac{\bigg((a + \sin( \pi/8) (1 - \sqrt{2} a)\bigg)\ket{111}+\frac{a}{\sqrt{2}} \ket{100}}{\sqrt{N}}
\end{align}
where,
\begin{align} 
	N =&\bigg((\frac{a}{\sqrt{2}} + \cos( \pi/8) (1 - \sqrt{2} a)\bigg)^2\nonumber\\
	&+\bigg((a + \sin( \pi/8) (1 - \sqrt{2} a)\bigg)^2+\frac{a^2}{2}
\end{align}
For the above form of the state, for any value of the parameter ‘a’, the measures GMC ($C_{GMC}$) and CF ($F_{123}$) can be calculated. It can thus be easily seen that for the representative examples of the pairs of values for the parameter ‘a’ given in various rows of the two Tables, viz. Table.\ref{tab1} and Table.\ref{tab:t2}, the measures $C_{GMC}$ and $F_{123}$ are not equivalent for the purpose of inferring which one of the given pair of states is more entangled than the other one. Furthermore, non-monotonicity between $C_{GMC}$ and $F_{123}$ has been illustrated in Fig.\ref{fig:8}.

\begin{figure}
	\includegraphics[scale=.32]{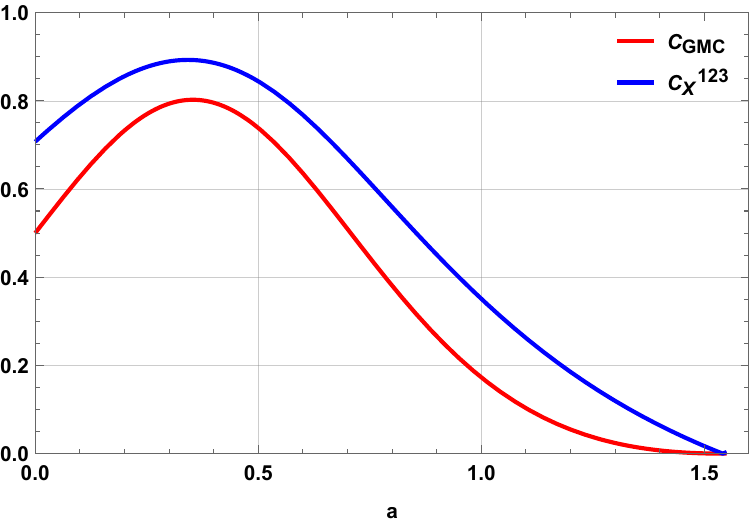}
	\includegraphics[scale=.23]{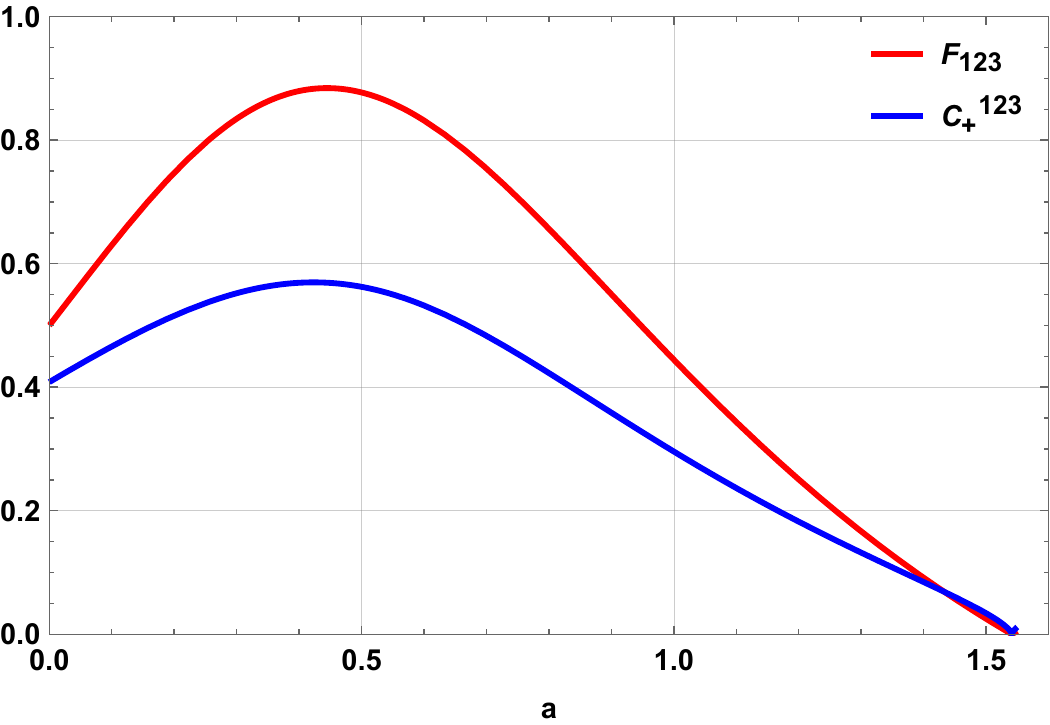}
	\caption{From the plot we can see that  $C^{123}_{X}$ is monotonically related to GMC and $C^{123}_{+}$ is monotonically related to triangle measure.  $C^{123}_{X}$ and $C_{GMC}$ has the same functional behavior with $a$, so if we know $C^{123}_{X}$ then we can calculate  $C_{GMC}$ . Similarly,  if we know $C^{123}_{+}$ then we can calculate  $F_{123}$.}
	\label{fig:9}
\end{figure}

Now, an important point to be noted is that one can also evaluate PCC for the above mentioned state given by Eq. \eqref{eqx0}. More specifically, by evaluating PCC with respect to the choice of the observable X for each of the subsystems, i.e., $C_{X}^{123}$ it can be checked that  $C_{X}^{123}$  is monotonically related to CGMC (Fig.\ref{fig:9}a). Further, by evaluating PCC with respect to the choice of the observable $\ket{+}\bra{+}$ for each of the subsystems, i.e.,  $C_{+}^{123}$  it can be seen that $C_{+}^{123}$ is monotonically related to $F_{123}$. Therefore, by empirically determining the Pearson Correlators $C_{X}^{123}$ and  $C_{+}^{123}$ for the respective members of a pair of states corresponding to any pair of values of the parameter ‘a’ in any of the rows of the Tables I and II, it will be possible to determine the genuine entanglement measures $C_{GMC}$ and $F_{123}$ for both the states, thereby showing their non-equivalence as illustrated in the Tables \ref{tab1} and \ref{tab:t2}. A highlight of this scheme is that, given the multiple choices available for  preparing the state given by Eq.\eqref{eqx0} with suitable values of the parameter ‘a’ (Tables \ref{tab1} and \ref{tab:t2}), there is considerable flexibility in exploring the experimental realizability of the scheme proposed here.”



\begin{widetext}\begin{center}
\begin{table}[ht] 
	\begin{tabular}
		{ |C{2.7cm}|C{2.7cm}|C{2.7cm}|C{2.7cm}|C{2.7cm}|C{2.7cm}|  }
		\hline
		\multicolumn{6}{|c|}{Different pairs of “a” for which we get “non-equivalence”} \\
		\hline
		$a_{1}$& $a_{2}$& $ C_{GMC}(a_{1})$ & $ C_{GMC}(a_{2})$&$F_{123}(a_{1})$&$F_{123}(a_{2})$  \\
		\hline
		0&   0.80  & 0.50 &0.38&0.50&0.66\\
		\hline
		0.10& 0.70&0.63 &0.51&0.63&0.75\\
		\hline
		0.20 & 0.60  & 0.73&0.64&0.75&0.83\\
		\hline
		0.25 & 0.55  & 0.77&0.69&0.80&0.86\\
		\hline
	\end{tabular}
	\caption{To highlight the non-equivalence let's focus on the first row which is $a_{1}=0$ and $a_{2}=0.80$. Here the measure $C_{GMC}$ says the first state is more entangled than the second state but  $F_{123}$ says the second state is more entangled than the first state, so we can clearly see the non-equivalence between these two measures.}
 \label{tab1}
\end{table}
\begin{table*}[h!]
	
	\begin{tabular}{ |C{2.7cm}|C{2.7cm}|C{2.7cm}|C{2.7cm}|C{2.7cm}|C{2.7cm}|   }
		\hline
		\multicolumn{6}{|c|}{Different pairs of “a” for which we get “non-equivalence”} \\
		\hline
		$a_{1}$& $a_{2}$& $ C_{GMC}(a_{1})$ & $ C_{GMC}(a_{2})$&$F_{123}(a_{1})$&$F_{123}(a_{2})$  \\
		\hline
		0&   0.71  & 0.50 &0.50&0.50&0.75\\
		\hline
		0.20& 0.51&0.73 &0.73&0.75&0.87\\
		\hline
		0 & 0.95  & 0.50&0.22&0.50&0.50\\
		\hline
		0.39 & 0.50  & 0.80&0.74&0.88&0.88\\
		\hline
	\end{tabular}
	\caption{To highlight the non-equivalence let's focus on the first row which is $a_{1}=0$ and $a_{2}=0.871$. Here the measure $C_{GMC}$ says both of the states have the same entanglement but  $F_{123}$ says the second state is more entangled than the first state. }
 \label{tab:t2}
\end{table*}
\end{center}
\end{widetext}

\subsubsection{Sudden Death of Tripartite Entanglement } \label{sec7}
If we start with an entangled state and evolve it through a thermal bath then it may happen that the entanglement decays down to zero at some finite time which is known as the sudden death of entanglement. This phenomenon has been studied extensively in the bipartite system \cite{PhysRevLett.93.140404,PhysRevA.69.052105,PhysRevA.73.040305}. Recently, entanglement sudden death (ESD) for tripartite system has been demonstrated using CF and GMC in \cite{2022arXiv221001854X}. It was shown that if we start from the generalized $\ket{GHZ}$ state and evolve it through a Markovian bath then the system will show sudden death of entanglement. Let's write the  generalized $\ket{GHZ}$ state in the following form
\begin{equation} \label{ghzsh}
	\begin{split}
		\ket{GHZ} = \sqrt{1-y}\ket{000}+\sqrt{y}\ket{111}\;,
	\end{split}
\end{equation}
where  $y\in[0,1]$ .
Now if our initial state has the above form then after time t (by evolving through a local amplitude damping
Markovian channel) the GMC will be \cite{2012PhRvA..86f2303H,2022arXiv221001854X}
\begin{equation} \label{eqsdgmc}
	\begin{split}
		C_{GMC} = \text{Max} \;   [0,\; |2 \sqrt{y(1-y)}|\; p^3(t)(1-6y q^3(t))]\;,
	\end{split}
\end{equation}
where $p(t)=\sqrt{e^{-t/\tau}} $ ,  $q(t)=\sqrt{1-e^{-t/\tau}}$ and $1/\tau$ is the damping rate. It has been shown in  \cite{2022arXiv221001854X} that the sudden death of entanglement occurs at the time
when the following equality is satisfied
\begin{align}\label{sdtr}
	\sqrt{(1/y-1)}=3 (1-e^{-t/\tau})^{3/2}\;.
\end{align}
We now focus on PCC for the same state given by Eq.\eqref{ghzsh}. We have calculated PCC for this state using the choice of the observable $ \ket{+}\bra{+}$ for each of the subsystems and it is given by”
\begin{equation}\label{eqsdpcc}
	C^{123}_{+}=\frac{2 e^{-3t/2\tau}}{\sqrt{3}} \sqrt{y(1 - y) }
\end{equation}
Now, note that using Eq.\eqref{eqsdgmc} and  Eq.\eqref{eqsdpcc}, one can obtain the following analytical relation between $C_{GMC}$ and $C_{+}^{123}$
\begin{align}\label{eqspccgme}
	C_{GMC}=&\frac{\sqrt{3} C^{123}_{+} }{\sqrt{y(1 - y) }}\bigg[\sqrt{y(1 - y)}\nonumber\\
	&\;\;\;\;\;\;\;\;\;\;\;\;\;\;\;\;-3y\bigg(1- \big(\frac{\sqrt{3} C^{123}_{+} }{2\sqrt{y(1 - y) }}\big)^{2/3}  \bigg)^{3/2}\bigg]\;.
\end{align}
\begin{figure}
	\includegraphics[scale=.65]{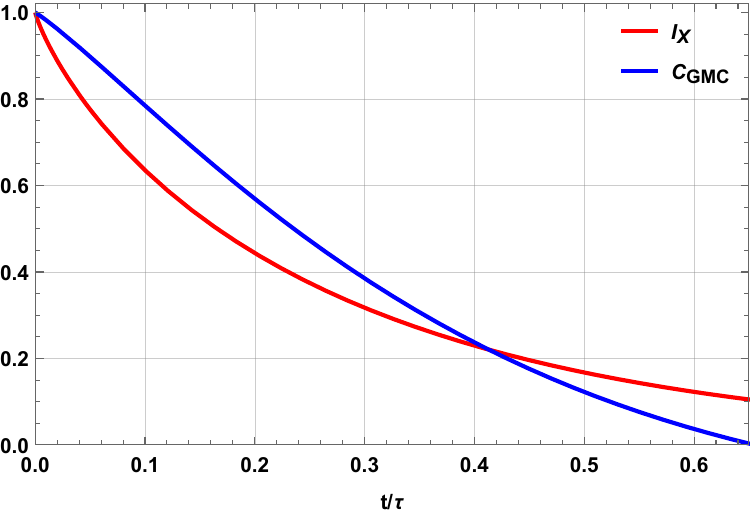}
	\caption{We have plotted $C_{GMC}$ and $I_{X}$ with $t/ \tau$ at $y=0.5$. From the plot, it is clear that they are monotonic i.e. they have the same functional behavior. Hence, using $I_{X}$ we can experimentally show the sudden death of entanglement. }
	\label{fig:13.0}
\end{figure}

\begin{figure}
\includegraphics[scale=.65]{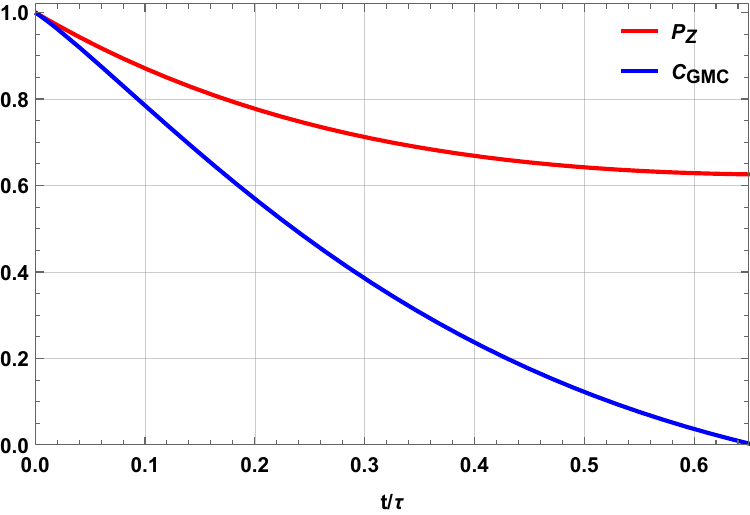}
	\caption{We have plotted $C_{GMC}$ and $P_{Z}$ with $t/ \tau$ at $y=0.5$. From the plot, it is clear that they are monotonic i.e. they have the same functional behavior. Hence, using $P_{Z}$ we can experimentally show the sudden death of entanglement. }
	\label{fig:13.00}
\end{figure}
By empirically determining  $C_{+}^{123}$ at appropriate instants while the initial state \eqref{ghzsh} evolves in the relevant damping channel, and using the above relation \eqref{eqspccgme} to evaluate the corresponding  $C_{GMC}$, one can therefore empirically verify the prediction of ESD occurring at the instant fixed by Eq.\eqref{sdtr} corresponding to a given value of the parameter y of the initially prepared state of the form specified by Eq. \eqref{ghzsh}.

Next, we observe that it is possible to empirically verify the predicted ESD for the initial state \eqref{ghzsh} by also using Mutual Information (MI) with respect to the $\sigma_{X}\times\sigma_{X}\times\sigma_{X}$ basis ($I_{X}$) and Mutual Predictability (MP) with respect to the $\sigma_{Z}\times\sigma_{Z}\times\sigma_{Z}$ basis ($P_{Z}$).
In these cases, for the time evolution of the state \eqref{ghzsh} in the same damping channel as considered earlier, we show by numerical computation the monotonic relationship between $C_{GMC}$ and  $I_{X}$,  $P_{Z}$ respectively, illustrated by the plots in the Figs. Fig.\ref{fig:13.0} and Fig.\ref{fig:13.00} respectively, taking the value of the state parameter $y=0.5$. Therefore, by empirically determining $I_{X}$ or $P_{Z}$ it is also possible to empirically verify the prediction of ESD according to Eq. \eqref{sdtr}.


\section{Discussion} \label{sec9}
Recently, statistical correlators got significant attention for characterizing the bipartite entanglement. However, till now no work has been done in the direction of characterizing tripartite entanglement using statistical correlators. In this paper, we start by defining statistical correlators in the multipartite system. Our definitions are consistent with the fact that whenever these tripartite states are product states or bi-separable, the measure should be zero.
Using this definition,  we have calculated statistical correlators for a few classes of pure tripartite entangled states.  Interestingly, we find 
monotonic relations between the statistical correlators and the genuine measures of tripartite entanglement for those pure tripartite states. 
These monotonic relations can be used to find the genuine measures of tripartite entanglement experimentally.

 In Ref.\cite{triangle}, Xie and Eberly have shown that the genuine entanglement measures, CF and GMC, are inequivalent in the sense of not being monotonic with respect to each other. However, to illustrate this inequivalence, they have taken only one particular example.
But in our work, we have explicitly shown that this is not a special example for which such inequivalence can be noticed.  In fact, it is possible to construct quite a number of pairs of states 
 for which we can get such inequivalence between GMC and CF. Also,
we have provided a scheme to empirically detect inequivalence between different genuine measures of tripartite entanglement using statistical correlators. 

 The possibility of tripartite entanglement being used for demonstrating the entanglement sudden death (ESD) has been revealed recently \cite{2022arXiv221001854X}, crucially by using CF and GMC. It is thus important to investigate experimental aspect of the sudden death of tripartite entanglement. In this paper, we have discussed how one can use statistical correlators to show the sudden death of tripartite entanglement empirically.  


One of the important future directions will be to investigate the relation between statistical correlators and genuine measures for other classes of mixed tripartite states. However, calculating genuine measures of tripartite entanglement requires convex-roof construction which is in general very challenging \cite{xie2022estimating, doi:10.1080/00107514.2022.2104425}. 
In \cite{2022arXiv221001854X}, very recently for a mixture of $|GHZ\rangle$ and  $|W\rangle$ state Concurrence fill was computed. Interestingly, the calculation of PCC or MI is relatively easy for the mixed state. The basic reason is in PCC or MI calculation, we calculate the expectation value which is independent of the basis we choose to write the mixed state and hence we do not require any convex roof construction. In Appendix. \ref{sec8}, we consider mixed state in detail. In particular, we compute PCC and MI for a mixture of GHZ and W state. We observe a  monotonic relation between genuine tripartite measure CF and statistical correlator MI. This gives us a way to determine entanglement experimentally for a mixed tripartite state.
See Appendix \ref{sec8} for more details. 


Global Measure of Entanglement($ G_{123}$) and  Tangle Measure of Entanglement($ \tau_{123}$) are two extensively used measures of tripartite entaglement  \cite{gm1,gm2, Singh:20, PhysRevA.61.052306}. 
However, these are not genuine measures of tripartite entanglement \cite{triangle}.  In Appendix \ref{ngm}, we review these two measures and investigate whether there exists a monotonic relation between the statistical correlators and these measures. More precisely, we got monotonic relation between $ G_{123}$, $ \tau_{123}$ and PCC for the generalized GHZ state but did not get monotonic relation between $ G_{123}$, $ \tau_{123}$ and PCC for the generalized W state.

Finally, we note that, very recently, Xie et al. introduced a genuine measure of entanglement for a four-qubit system in \cite{xie2023unraveling}. It will be interesting to investigate the relation between this genuine measure and statistical correlators by generalizing the statistical correlators for the quadripartite system.



 \section*{Acknowledgements} 
The work of SJ is supported by Ramanujan Fellowship. SK acknowledges the CSIR fellowship with Grant Number 09/0936(11643)/2021-EMR-I.  US would like to acknowledge partial support provided by the Ministry of Electronics and Information Technology (MeitY), Government of India under the grant for Center for Excellence in Quantum Technologies with Ref. No. 4(7)/2020 – ITEA and QuEST-DST project Q-97 of the Govt. of India. The authors would also like to thank the people of India for their steady support in basic research.

\bibliographystyle{apsrev4-2}
\bibliography{abc.bib}

\appendix
\section{Non-genuine measures of tripartite entanglement}\label{ngm}
In this appendix, we are going to review two well known but non-genuine measures of tripartite entanglement, the Global Measure of Entanglement($ G_{123}$) and the Tangle Measure of Entanglement($ \tau_{123}$).
According to \cite{triangle}, the Global Measure of Entanglement is not a genuine measure because it is non-zero for some product and biseparable states which violates the condition (a) discussed in  \ref{sec3}. 
Whereas,  the Tangle Measure of Entanglement is not a genuine measure because it is zero for some nonbiseparable
states (GHZ class and W class in the three-qubit case) which violates condition (b) discussed in  \ref{sec3}.

\subsection{Global Measure of Entanglement($ G_{123}$)}
 The global measure of entanglement
for a generalised three-qubit state $\rho$ defined as \cite{gm1,gm2,Singh:20},
\begin{equation}\label{eqg}
    G_{123}(\rho)= 2\Big[1-\frac{P^2_{1(23)}+P^2_{2(13)}+P^2_{3(12)}}{3}\Big]
\end{equation}
where, $P_{i(jk)}=\sqrt{Tr_{jk}[Tr_{i}\rho]^2}$ with $i = 1, 2, 3$ and $jk =
23, 13, 12$ respectively.
\subsection*{Examples:}
\subsubsection{Generalized $\ket{GHZ}$ state}
Using equation \eqref{eqg}, we can show that the global measure of entanglement($G_{123}$) for the generalized $\ket{GHZ}$ state
 is
 \begin{equation}\label{eqghzg}
     G_{123}=4a^2b^2
 \end{equation}
By comparing Eq.\eqref{eqcghz} and Eq.\eqref{eqghzg} we got the following analytical relation between $C^{123}_{X}$ and the global measure of entanglement($G_{123}$),
\begin{equation}\label{eqcg}
     C^{123}_{X}=\sqrt{G_{123}}
\end{equation}
For this state, $G_{123}$ is monotonic with $C^{123}_{X}$.
\subsubsection{Generalized $\ket{W}$ state}
\begin{figure}
    \includegraphics[scale=.7]{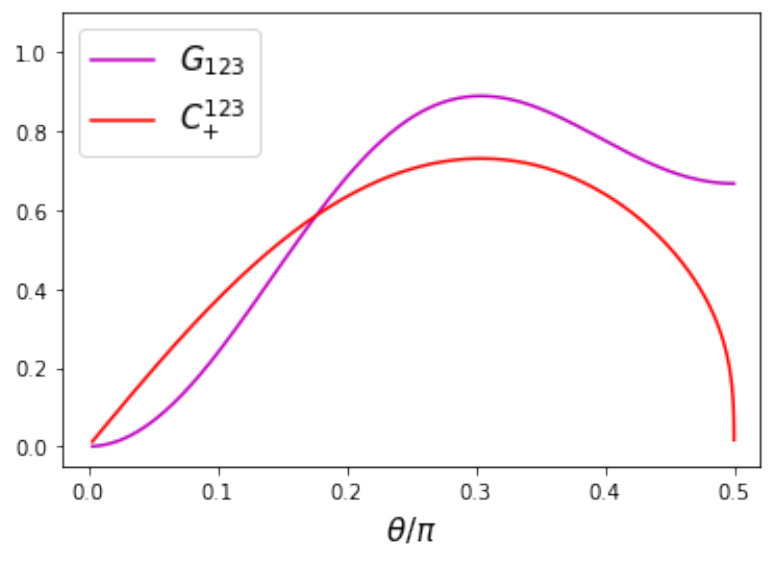}
    \caption{From the plot, we can see that $G_{123}$ is not decaying to zero when $\theta$ is going to $\pi/2$ (bi-product state), so the global measure of entanglement($G_{123}$) is not a good measure of entanglement for generalised $\ket{W}$ state.}
    \label{fig:100}
\end{figure}

The global measure of entanglement($G_{123}$) for the generalized $\ket{W}$ state 
 is given by,
\begin{equation}\label{eqwg}
G_{123}=\frac{2}{3}\big[\sin^2{\theta}(1+3\cos^2{\theta})]
\end{equation}
We have explicitly calculated $C^{123}_{+}$ for the generalized $\ket{W}$ state in Eq.\eqref{eqwc1}. We have plotted $C^{123}_{+}$ and $G_{123}$  with the free parameter $\theta$ in Fig.\ref{fig:100}. From the plot, it is clear that $G_{123}$ is not monotonic with $C^{123}_{+}$.
\subsection{Tangle Measure of Entanglement($ \tau_{123}$)}
In \cite{PhysRevA.61.052306},  Coffman et al. introduced the Tangle measure of entanglement, and
it was calculated for both generalized $\ket{GHZ}$ state and generalized $\ket{W}$ state in  \cite{Singh:20}. For generalized $\ket{GHZ}$ state, they got
 \begin{equation}\label{eqghzt1}
     \tau_{123}=4a^2b^2\;.
 \end{equation}
By comparing Eq.\eqref{eqcghz} and Eq.\eqref{eqghzt1} we got the following analytical relation between $C^{123}_{X}$ and the tangle measure of entanglement($\tau_{123}$),
\begin{equation}\label{eqcg}
	C^{123}_{X}=\sqrt{\tau_{123}}
\end{equation}
For this state, $\tau_{123}$ is monotonic with $C^{123}_{X}$.

However, for the generalized $\ket{W}$ state, the tangle measure is zero.

\section{Extension of Maccone conjecture in the tripartite system: PCC to detect tripartite entanglement }\label{tmc}
For the bipartite system, PCC can be used to detect bipartite entanglement \cite{macc,sinha}.
In this appendix, we are going to investigate whether the Maccone conjecture \cite{macc} is valid for the tripartite case or not. To do that let us first assume $A^1_{1}$, $A^1_{2}$ are two  complementary observables  of subsystem-1(1st qubit) and  similarly $B^{23}_{1}$, $B^{23}_{2}$ are two  complementary observables  of subsystem-(23). Now let's define the following quantities,
\begin{align}
C^{1-23}_{A_{1}B_{1}}= |C_{A^1_{1}B^{23}_{1}}|,\;\;  C^{1-23}_{A_{2}B_{2}}= |C_{A^1_{2}B^{23}_{2}}|\;,
\end{align}
 where $C_{A^1_{i}B^{23}_{i}}$ is the pearson correlation coefficient defined in equation (\ref{eqpcc}) . Similarly, for the other two combinations, we can define the following quantities- $C^{2-13}_{A_{i}B_{i}}$ and $C^{3-12}_{A_{i}B_{i}}$. Let us now consider the following quantity (sum of PCCs),
\begin{equation}\label{eqapcc}
	C^{123}=\sum_{i=1}^{2}(C^{1-23}_{A_{i}B_{i}}C^{2-13}_{A_{i}B_{i}}C^{3-12}_{A_{i}B_{i}})^{\frac{1}{3}}
\end{equation}
We have calculated the above quantity for generalized $\ket{GHZ}$ and for generalized $\ket{W}$ state to verify the conjecture.

\subsection*{Examples:}
\subsubsection{ Generalized $\ket{GHZ}$ state}
The generalized  $\ket{GHZ}$ state has the following form,
\begin{equation} 
	\begin{split}
		\ket{GHZ} = a\ket{000}+b\ket{111}
	\end{split}
\end{equation}
To calculate $C^{123}$, we have taken the following observables,
\begin{align}
	&A^1_{1}=A^2_{1}=A^3_{1}=\sigma_{x},B^{23}_{1}=B^{13}_{1}=B^{12}_{1}=\sigma_{x} \otimes \sigma_{x}\nonumber\\ &A^1_{2}=A^2_{2}=A^3_{2}=\ket{1}\bra{1}, B^{23}_{2}=B^{13}_{2}=B^{12}_{2}=\ket{1}\bra{1} \otimes \ket{1}\bra{1}  
\end{align}
Using equation \eqref{eqapcc}, we can show that $C^{123}$ for this state is given by,
\begin{equation}\label{eqmcghz}
	C^{123}=1+2|ab|
\end{equation}
We can clearly see that $C^{123}$ is always greater than $1$ except for $a=0$ or $b=0$ which correspond to the separable state. So for the generalized GHZ state, the conjecture is true. 
\subsubsection{Generalized $\ket{W}$ state}
The generalized $\ket{W}$ state has the following form,
\begin{equation} 
	\begin{split}
		\ket{W} = \cos{\theta}\ket{100}+\frac{\sin{\theta}}{\sqrt{2}}\ket{010}+\frac{\sin{\theta}}{\sqrt{2}}\ket{001}
	\end{split}
\end{equation}
To calculate $C^{123}$, let's take the following observables,
\begin{align}
	&A^1_{1}=A^2_{1}=A^3_{1}=\ket{0}\bra{0}, B^{23}_{1}=B^{13}_{1}=B^{12}_{1}=\ket{0}\bra{0} \otimes \ket{0}\bra{0} \nonumber\\ &A^1_{2}=A^2_{2}=A^3_{2}=\ket{+}\bra{+}, B^{23}_{2}=B^{13}_{2}=B^{12}_{2}=\ket{+}\bra{+} \otimes \ket{+}\bra{+}
\end{align}
Using equation \eqref{eqapcc}, we can show that $C^{123}$ for this state is given by,
\begin{align}\label{eqmcw}
&	C^{123}=1\nonumber\\
	&+\bigg[\bigg(\frac{\sqrt{2}\sin{2\theta}}{\sqrt{(1+\sin^2{\theta})(2+\cos^2{\theta})}}\bigg).\bigg(\frac{\sin{2\theta}+\sqrt{2}\sin^2{\theta}}{\sqrt{6+2\sqrt{2}\sin{2\theta}-\sin^2{2\theta}}}\bigg)^2\bigg]^{\frac{1}{3}}   
\end{align}\\
For the generalized W state also, $C^{123}$ is always greater than $1$ except for $\theta=0$ or $\theta=\pi/2$ which correspond to the bi-separable state. So the conjecture is true for the generalized W state also. Here, we have checked the conjecture in a few examples but it would be nice to check it extensively to have a comprehensive study of the conjecture. 
It may happen that for the tripartite case, the bound can be made stronger.

\section{PCC and MI for mixed state } \label{sec8}
 In this Appendix, we want to investigate if it is possible to extend the analysis of pure state done in the main text for mixed state. As pointed out in the discussion section, calculating genuine measures like GMC and CF
for the mixed state is difficult but we can easily calculate PCC and MI.  
\subsection*{Calculation of PCC and MI for mixed state}
Calculating tripartite entanglement such as CF or GMC is very difficult for mixed states. Calculation of these measures in mixed states requires convex roof construction. In \cite{2022arXiv221001854X}, very recently for a mixture of $|GHZ\rangle$ and  $|W\rangle$ state, CF was computed. Interestingly, the calculation of PCC or MI is relatively easy for the mixed state. The basic reason is in PCC or MI calculation, we calculate the expectation value which is independent of the basis we choose to write the mixed state and hence we do not require any convex roof construction. 

In the rest of the section, we calculate PCC and MI for a mixture of $|GHZ\rangle$ and  $|W\rangle$ state and compare them with the CF.
\begin{figure}[h!]
\includegraphics[scale=.65]{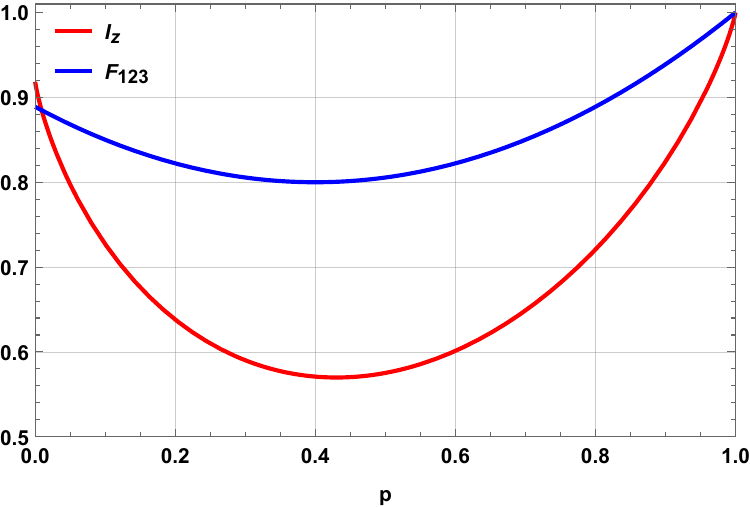}
	\caption{We have plotted $F_{123}$ and $I_{z}$ with $p$. From the plot, it is clear that they are monotonic.}
	\label{fig:13}
\end{figure}
\subsubsection*{ Mixture of $\ket{GHZ}$ and $\ket{W}$ state:}
Let's take the following mixed state
\begin{equation} \label{eqm2}
	\begin{split}
		\rho(p) =p \ket{GHZ}\bra{GHZ}+(1-p)
		\ket{W}\bra{W}\;.
	\end{split}
\end{equation}
Here we first calculate $C^{123}_{AB}$ with respect to the $\ket{0}\bra{0} $ operator i.e.
\begin{align}\label{eqm2o1}
	&A^1_{1}=A^2_{1}=A^3_{1}=\ket{0}\bra{0}, B^{23}_{1}=B^{13}_{1}=B^{12}_{1}=\ket{0}\bra{0} \otimes \ket{0}\bra{0} 
\end{align}
Using equation \eqref{eqspcc}, we can show that $C^{123}_{AB}$ for this state is given by
\begin{equation}\label{eqm2c0}
	C^{123}_{\ket{0}}=\frac{p^2+16p-8}{(2+p)(4-p)}\;.
\end{equation}
Similarly, $C^{123}_{AB}$ with respect to the $\ket{+}\bra{+} $ operator is
\begin{equation}\label{eqm2c+}
	C^{123}_{+}=\frac{4-p}{\sqrt{(5-2p)(7+2p)}}\;.
\end{equation}

MI ($I_{Z}$) for this mixed state with respect to the $\sigma_{z}$ basis is given by 
\begin{align}
	I_{z}=&\frac{2(1-p)}{3}\log_{2}\bigg(\frac{6}{4-p}\bigg)+\bigg(\frac{1-p}{3}\bigg)\log_{2}\bigg(\frac{12(1-p)}{(2+p)^2}\bigg)\nonumber\\
	&+\frac{p}{2}\;\bigg[\log_{2}\bigg(\frac{6}{2+p}\bigg)+\log_{2}\bigg(\frac{18p}{8+2p-p^2}\bigg)\bigg]\;.
\end{align}
In Ref.\cite{2022arXiv221001854X}, the authors calculated the CF ($F_{123}$) for the mixture of $\ket{GHZ}$ and $\ket{W}$ state (defined in Eq.\eqref{eqm2})  using convex roof. They got the following analytical expression for $F_{123}$
\begin{equation}
	F_{123}=\frac{5 p^2-4p+8}{9}\;.
\end{equation}
\subsection*{Relation between PCC, MI and Concurrence Fill for the mixture of $\ket{GHZ}$ and $\ket{W}$ state }
We have calculated PCC  with respect to the  $\ket{0}\bra{0} $ and $\ket{+}\bra{+} $ operators in Eq.\eqref{eqm2c0} and Eq.\eqref{eqm2c+} respectively. We have also calculated MI for the same state.  We got a monotonic relation between $F_{123}$ and MI with respect to the $\sigma_{z} $ operator, $I_{z}$, but we didn't get a monotonic relation between  $F_{123}$ and PCC. However, one can calculate PCC using some other basis to search for a monotonic relation between PCC and $F_{123}$. 
In Fig.\ref{fig:13} we have plotted  $F_{123}$ and $I_{z}$ with the free parameter $p$.
From the plot, it is clear that they are monotonic.


\end{document}